\documentclass{aip-cp}
\usepackage{amsfonts,amssymb,amsbsy,amscd}
\usepackage[numbers]{natbib}
\usepackage{rotating}
\usepackage{graphicx}
 \newtheorem{theorem}{Theorem}
\newtheorem{example}{Example}
 \newtheorem{lemma}{Lemma}
 \newtheorem{remark}{Remark}
 \newtheorem{corollary}{Corollary}
\def\eqref{(\ref)}

\begin{document}

\title{Mean field game equations with underlying jump-diffusion process}

\author[aff1]
{Olga Rozanova}
\eaddress{Corresponding author: rozanova@mech.math.msu.su}
\author[aff1]{Ilnar Manapov}
\affil[aff1]{Moscow State University, Moscow, 119991, Russian Federation.}


\maketitle

\begin{abstract}We consider a couple of integrodifferential PDEs arising from a stochastic Markovian control problem subjected to initial-terminal conditions. These equations correspond to the MFG system for a controlled jump-diffusion process.
We prove that for a specific choice of the control function the
expectation of the jump-diffusion process can be found explicitly.
 The study is an extension of similar results known for the pure diffusion process. As an example, we show how this can be applied to the problem of investors evaluating the trend of an asset when choosing an optimal portfolio.
\end{abstract}



\section{1. CONCEPT OF MEAN FIELD GAMES (MFG) }

MFG theory studies models with a large number of small components (agents) that interact with each other achieving their individual objectives. 
The term "mean field" means that the strategy of each agent to achieve the maximum of its individual utility directly depends on the average distribution of influences of other agents and does not depend on the initial configuration of the system. The mean field theory is well known in statistical physics, but similar concepts related to active objects were formulated only in the last decade.  These concepts, together with the optimal control theory, have made it possible to study models in economics and sociology. Now MFG are widely used  in areas requiring analysis of differential games with a large number of participants (e.g. \cite{Caines}, \cite{Gomes}, \cite{GLL}).

\subsection{1.1 \,A heuristic derivation of MFG equations }

Here we give a very sketchy derivation of the basic equations,
referring for a more thorough exposition, for example, to
\cite{Gomes}. Suppose that the averaged behavior of the agents is
described by  a real-valued stochastic process
\begin{equation}\label{X}
dX_s = \alpha_s\, d s + \delta\, d W_s + \lambda d \Gamma_s, \quad X_t=x,
\end{equation}
  $x\in \mathbb R$  is a point in the space of the states,
  $0\le t\le s\le T$,
$W_s$ is a standard  Brownian motion, $\Gamma_s$ is a  Poisson process (jumps),
 $\delta\ge 0$,  $\lambda\ge 0$
 are constants,
 $\alpha_s=\alpha(s,X_s)$ is an admissible Markov control (choosing the value from a given Borel set $U\subset \mathbb R $ at any time $ s $, one can control the process
$X_s$).
 Fixing
$x$, we mark trajectory for a specific agent.

We denote the probability density of process  (\ref{X}) 
 as $m(t,x)$,  $\int_{\mathbb R}m(t,x)=1$.

 The problem of control is to define maximum over all the admissible controls $\alpha_t$
$$J(t,x;\alpha_t) = {\mathbb E}\Bigg[
\int_t^T F(s,X_s;\alpha_s)ds + K(X_T) \Bigg], $$
 $F:{\mathbb
R}_+\times {\mathbb R}\times U\to \mathbb R$ and $K:{\mathbb R}\to
\mathbb R$  are prescribed continuous functions,
 the process  $X_s$ obeys (\ref{X}).

{The payoff  function} $\Phi: {\mathbb R}_+\times
\mathbb R$ is considered as
$$\Phi (t,x)=\sup\limits_{\alpha\in U}\,J(t,x;\alpha). $$

The Hamilton-Jacobi-Bellman equation which allows to solve the problem above is (see \cite{Oksendal1} )
\begin{equation}
\label{HJ} \sup\limits_{\alpha\in U}\Bigg( F(t,x;\alpha) +
\Big(\mathcal{L}^{\alpha}\Phi\Big)(t,x) \Bigg) =
0,
\end{equation}
where
\begin{eqnarray}
\Big(\mathcal{L}^{\alpha}\Phi\Big) =
\frac{\partial \Phi}{\partial t} + \alpha\frac{\partial
\Phi}{\partial x}
 + \frac{\delta^2}{2}\frac{\partial^2 \Phi}{\partial x^2}
  + \lambda \left(\int\limits_{\mathbb R} \Phi(x-z)p(z) dz - \Phi(x) \right),\nonumber
\end{eqnarray}
where $p(z)$ is a distribution density of jumps, $\int\limits_{\mathbb R}\,p(z) dz=1$,
with the terminal condition
\begin{eqnarray}\nonumber
\label{(4)} \Phi(T,x) = K(x(T)).\nonumber
\end{eqnarray}

We consider  a particular case
\begin{equation}\nonumber
\label{(1)}F(s,X_s;\alpha)= -\frac{\alpha^2}{2} + g(m(s,X_s),s,X_s).
\end{equation}
We assume that each agent
\begin{itemize}
\item receives a penalty for changing its position in the phase space;
 \item seeks to maximize its individual utility function, based on the fact that he knows only the distribution of the other agents.
\end{itemize}

The choice of the function  $g$ depends on the type of problem.

Based on the specific form of the quality function  $F$, we obtain from
(\ref{HJ})
\begin{eqnarray}\nonumber
    \partial_t \Phi + \sup_{\alpha} \left( \alpha \partial_x \Phi  - \frac{1}{2} \alpha^2\right) + g(m,t,x)
  +  \frac{\delta^2}{2} \partial^2_{xx} \Phi + \lambda \left( \int\limits_{\mathbb R}  \Phi(x-z) p(z) dz - \Phi(x)  \right) = 0.
\end{eqnarray}

 We get
 {the initial-terminal problem for
coupled Hamilton-Jacobi-Bellman (HJB) and Kolmogorov-Feller-Fokker-Planck (KFFP) equations}:
\begin{eqnarray}
        & \partial_t \Phi + \frac{1}{2} (\partial_x \Phi)^2 + \frac{\delta^2}{2} \partial^2_{xx} \Phi + \lambda \left( \int_{0}^{x} \Phi(x-z) p(z)  dz - \Phi(x)  \right) = -g, \label{HJB}\\
        & \partial_t m + \partial_x(m\partial_x \Phi) - \frac{\delta^2}{2} \partial^2_{xx}m  - \lambda \left( \int_{0}^{x} m(x-z) p(z)  dz - m(x)  \right) = 0, \label{KFP}\\
        & m(0,x) = m_0(x),\label{BC_KFP}\\ 
        &\Phi(x,T) = K(x),\label{BC_HJB}
\end{eqnarray}
$x\in \mathbb R$, $t\in [0,T]$.

In what follows we denote the expectation of $X_t$ as ${\mathbb E}(t)$ and  assume that it exists.

\bigskip

The aim of this work is
\begin{itemize}
\item   To show that for some choice of $g$ and $K$ the expectation ${\mathbb E}(t)$ can be found analytically
by means to reduction  to a system of Riccati equations;
\item To give an example of economic problem ({the problem of forming an investor's opinion on an asset})  where  $g$ and $K$ have a form that allows to find a solution.
\end{itemize}

From a mathematical point of view, it is interesting to analyze the function $ g $, which depends on $ m $, because in this case the equations HJB and KFFP are linked. In this paper we consider a specific form of this dependence, $g(\mathbb E(t), t, x)$.

\section{2. PROPERTIES OF THE EXPECTATION OF $X_t$}\label{S2}
\subsection{2.1.\,Hamilton-Jacobi-Bellman equation}

We start from analysis of the Hamilton-Jacobi-Bellman equation (\ref{HJB}).
Assume
\begin{equation}\label{g} g = a(t) x^2 + b(t)x +
c(t), 
\end{equation}
where $a, b, c$  are continuous functions on $[0,T]$.

 We  look for a solution  in the form
\begin{equation}
\label{f2} \Phi = { A}(t)x^2 + { B}(t)x + { C}(t), \quad
\end{equation}
which imposes the terminal condition (\ref{BC_HJB}) as
\begin{equation}
\label{f2K} K(x)={ A}_T x^2+{ B}_T x +{ C}_T,
\end{equation}
with constants  ${ A}_T, { B}_T, { C}_T$.

\bigskip

The following lemma can be checked by a straightforward calculation.

 \begin{lemma}
 Assume that the second moment of the distribution of jumps $\int\limits_{\mathbb  R}\,z^2 p(z)\, dz $ is finite and $g$ has the form (\ref{g}). Then (\ref{HJB}) has a solution of the form (\ref{f2}). The coefficients $A(t)$, $B(t)$, $C(t)$ are uniquely defined by the terminal conditions  (\ref{f2K}).
 \end{lemma}

Proof. We substitute (\ref{f2}) to equation (\ref{HJB}) and get a polynomial of the second order with respect to $x$. Then we combine the coefficients at $x^2$, $x$ and $1$ and get a system of ODE equations for $A, B, C$, subject to the terminal conditions. The equation for $A$ splits from the rest of system and defines the entire dynamics. Equation for $B$ and $C$ are linear and contain $A$ in coefficients.
In particular, $A, B$ satisfy the system of ODEs with terminal conditions
\begin{eqnarray}
        \dot{A}+2 A^2 = - {a},  \quad
        \dot{B}+2 AB = - {2\lambda} \mathcal M \, A -{b},  \quad \mathcal M=\int\limits_{\mathbb R} \, z p(z)\,dz,
      \quad  {A(T) = A_T}, \quad
        B(T) = B_T. \label{AB}
         \end{eqnarray}
$\Box$

\begin{remark}
Coefficients $A(t)$ and $B(t)$ can blow up within the interval $(0,T)$.
\end{remark}

\begin{example}
As an example, we can consider two extreme cases of densities corresponding to jumps:
\begin{enumerate}
\item  jumps in positive direction only, $p(z)=p_+(z)$, $z\ge 0$,  where $\int\limits_0^\infty\,p_+(z)\, dz=1 $;

\item jumps when both directions are equal, $p(z)=\frac12 p_+(z)$, $z\ge 0$,  $p(z)=\frac12 p_+(-z)$, $z< 0$.
\end{enumerate}
In case 2, the mathematical expectation of the distribution of jumps $\mathcal M $ is zero.
\end{example}

\medskip

\subsection{2.1.\,Kolmogorov-Feller-Fokker-Planck  equation}
The second step is finding  the fundamental solution to KFFP equation (\ref{KFP}).
If $\Phi(t,x)$ is known, (\ref{KFP}) takes a more specific form,
\begin{eqnarray}\label{m1}
    \partial_t m + {( 2A \,x + B)} \partial_x m  + {2A} m - \frac{\delta^2}{2} \partial_{xx}^2 m - \lambda
    \left(   \int_{-\infty}^{\infty} m(t,x-z) p(z) dz - m  \right) = 0,
\end{eqnarray}
First we consider initial condition
\begin{equation}\label{FS}
m_0(x) = \delta(x)
\end{equation}
and denote the solution to (\ref{m1}), (\ref{FS})  as $\mathcal G(t,x)$.

We apply the Fourier transform $x\to \omega$ to obtain the characteristic function $\hat{m}=\hat{m}(t,\omega)$:
\begin{eqnarray}
 \partial_t \hat{m} - 2{A(t)}\omega \partial_\omega \hat{m} +  \left(\frac{\delta^2}{2}  \omega^2 + i B(t)\omega  +   (\hat{p}(\omega)-1) \lambda \right) \hat{m} = 0,
   \quad \hat{m} (0,\omega) = 1,\label{meq}
\end{eqnarray}
where it is assumed that $ A $ and $ B $ are known from the previous step.

The solution to (\ref{meq}) can be explicitly found, namely,
\begin{eqnarray}\label{mhat}
\hat \mathcal G (t,\omega)=
    {\exp}\left[-\int_{0}^{t}   \left(\frac{\delta^2}{2}
    {\mathcal R}^2 + i B(\eta)  {\mathcal R}  +   (\hat p({\mathcal R})-1)\lambda\right) d\eta\right],\quad   {\mathcal R}= {\mathcal R}(t,\eta,\omega)= \omega e^{2 R(t,\eta)}, \quad R(t,\eta)=\int\limits_\eta^t A(\tau) d\tau.
   \end{eqnarray}

\begin{theorem}
Let  $X_t$  be a random process satisfying (\ref{X}) with an initial probability density (\ref{BC_KFP}), $x \,m_0(x) \in L_1({\mathbb R)}$. If the coefficient $A(t)$ and $B(t)$, a  solution to (\ref{AB}), are such that $\hat \mathcal G (t,\omega)$ has a finite derivative with respect to $\omega$ for all $t\in (0,T)$, then
 the expectation of $X_t$ can be found as
  \begin{equation}\label{T1E}
    \mathbb{E}(t) =\int_{0}^{t}e^{2\int\limits_\eta^t A(\tau) d\tau} (B(\eta)+\lambda \mathcal M) \,d\eta + \mathbb{E}(0),\quad t\in [0,T].
\end{equation}
\end{theorem}
  Proof.
  Let   $X^0_t$  be a random process satisfying (\ref{X}), having initial probability density $m_0(x)=\delta(x)$. The solution to (\ref{m1}) in this case is the fundamental solution $\mathcal G(t,x)$.
  Then, by means of standard computations we get from  (\ref{mhat})
 \begin{equation}\label{E}
    \mathbb{E}(X^0_t) = i \frac{\partial \hat{\mathcal G}}{\partial \omega} (t,0) =\int_{0}^{t}e^{2 R(t,\eta)} \left(B(\eta)+{\lambda}{\mathcal M}\right) \,d\eta.
\end{equation}
The solution to (\ref{m1}) with arbitrary integrable data can be found as a convolution of $\mathcal G $ with $m_0(x)$, therefore
$ \hat m(t,\omega)= \hat \mathcal  G (t,\omega)\,\hat m_0(\omega)$,
$$ \mathbb{E}(X_t) = i \frac{\partial \hat{ m}}{\partial \omega} (t,0)=  i \frac{\partial \hat{\mathcal G}}{\partial \omega} (t,0) \hat{ m_0} (t,0)+ i \frac{\partial \hat{m_0}}{\partial \omega} (t,0) \hat{\mathcal G} (t,0).$$

Since $\hat{ m_0}(t,0)=\hat{\mathcal G} (t,0)=1$, we get (\ref{T1E}). $\Box$

\bigskip

\begin{theorem}
Let  $X_t$  be a random process satisfying (\ref{X}), having initial probability density $m_0(x)$ and $a(t)\equiv\rm const$. If the coefficients $A(t)$ and $B(t)$, a  solution to (\ref{AB}), are such that $\hat \mathcal G (t,\omega)$ has a finite derivative with respect to $\omega$ for all $t\in (0,T)$, then
   $\mathbb{E}(t)$ satisfies the following linear boundary value problem
   \begin{equation}
\label{EE} \mathbb{E}^{''}(t) + 2a\mathbb{E}(t)=-b(t), \quad \mathbb{E}(0)=x_0, \quad
    \mathbb{E}(T) =\int_{0}^{T}e^{2\int\limits_\eta^T A(\tau) d\tau} \left(B(\eta)+{\lambda}{\mathcal M}\right) \,d\eta + \mathbb{E}(0),\quad t\in [0,T],
\end{equation}
\end{theorem}
  Proof. We differentiate (\ref{E}) twice and take into account (\ref{AB}) to obtain
  \begin{equation}\nonumber
 \mathbb{E}^{''} =  B'+2A\,\left(B+\frac{\lambda}{k}\right)+2\,\int_{0}^{t}e^{2R(t,\eta)}(A'+2A^2) \left(B(\eta)+{\lambda}{\mathcal M}\right) \,d\eta=-2\,\int_{0}^{t}e^{2R(t,\eta)} a(\eta) \left(B(\eta)+{\lambda}{\mathcal M}\right) \,d\eta-b.
\end{equation}
For $a=\rm const$ it implies (\ref{EE}). $\Box$

\bigskip
\begin{corollary}\label{Cor}
If $a(t)=\rm const$ and $b(t)= b_0+b_1\mathbb E(t)+b_2\mathbb E'(t) $, $b_0, b_1, b_2=\rm const$,  then  (\ref{EE}) transforms into a linear second order ODE
  \begin{eqnarray}\nonumber
\mathbb{E}^{''} + b_2\mathbb{E}'+(2a+b_1)\mathbb{E}=-b_0,
\end{eqnarray}
which can be explicitly solved.
\end{corollary}

\medskip

\subsection{2.1. \, The behavior of $\,\mathbb E(t)\,$ for constant $a$ and $b$}
 As we can see from Corollary \ref{Cor}, in the case of the behavior of $\mathbb{E}(t)$ is very simple. Namely,
 \begin{itemize}
 \item for $a>0$ it oscillates near the constant $-\frac{b}{2a}$,
 \item for $a< 0$, it tends to reach $-\frac{b}{2a}$  outside  boundary layers near $t=0$ and  $t=T$ if $T$ is sufficiently large,
 \item for $a= 0$ it is a polynomial with respect to $t$, at most quadratic.
 \end{itemize}

For constant $a$ and $b$ the solution to (\ref{AB}) can be found explicitly.
For example, for $a>0$ we have
\begin{eqnarray}
  A(t) &=& \sqrt{\frac{a}{2}} \,\tan \theta(t,T),\qquad \theta(t,T)=\arctan \sqrt{\frac{2}{a}} A_T+\sqrt{2a}(T-t),\nonumber\\
  B(t) &=& -\frac{b}{\sqrt{2a}}\,\tan \theta(t,T)+\frac{bA_T+aB_T}{\sqrt{a(a+2A_T^2)}\cos \theta(t,T)}+\left(1-\frac{\sqrt{a}}{\sqrt{a+2A_T^2}\cos \theta(t,T)}\right){\lambda}{\mathcal M},\nonumber
  \end{eqnarray}
 Moreover, the solution with any terminal condition $A_T$ can exist only for
$T<\frac{1}{\sqrt{2a}}\left(\frac{\pi}{2}-{\arctan}\, \sqrt{\frac{2}{a}} A_T\right)$.
Therefore it is pretty amazing that the behavior of the expectation $\mathbb E$ is so simple and exists for all $t\in [0,T]$ for almost all $T$. It can be explained by the fact that $A$ and $B$ have singularity at the same points and in computations of $\mathbb E$ these singularities cancel.
The terminal value  $\mathbb E(T)$, found by formula from (\ref{EE}), is
\begin{eqnarray}\label{ET}
  \mathbb E(T) &=& -\frac{b}{2\sqrt{a(a+2A_T^2)}\cos \theta(0,T)}+\frac{bA_T+aB_T}{\sqrt{2a}(a+2A_T^2)}\tan \theta(0,T)+
  \left(A_T-\sqrt{\frac{a}{2}}\right) \frac{{\lambda}{\mathcal M}}{ a+2A_T^2},
  \end{eqnarray}
it is finite if $\theta(0,T)\ne \frac{\pi}{2}(1+2n)$, $n\in \mathbb Z$.

Also it is interesting to trace the influence of the noise to the behavior of $\mathbb E(t)$. It is easy to see that the Wiener process, in contrast to the Poisson process, has no influence. Though the "average" value of  $\mathbb E(t)$, $-\frac{b}{2a}$, does not depend of $\lambda$, the amplitude of oscillations depends but for non-symmetric jumps omly (see (\ref{ET})). The amplitude increases together with the strength of jumps and the expectation $\mathcal M$.

 For $a<0$ the result is similar:
\begin{eqnarray}
  A(t) &=& \sqrt{\frac{-a}{2}} \,\tanh \theta(t,T),\qquad \theta(t,T)={\rm arctanh} \sqrt{\frac{2}{-a}} A_T+\sqrt{-2a}(T-t),\nonumber\\
  B(t) &=& -\frac{b}{\sqrt{-2a}}\,\tanh \theta(t,T)+\frac{bA_T+aB_T}{\sqrt{-a(-a+2A_T^2)}\cosh \theta(t,T)}+\left(1-\frac{\sqrt{-a}}{\sqrt{-a+2A_T^2}\cosh \theta(t,T)}\right){\lambda}{\mathcal M},\nonumber
  \end{eqnarray}
\begin{eqnarray}\label{ET}
  \mathbb E(T) &=& -\frac{b}{2\sqrt{-a(-a+2A_T^2)}\cosh \theta(0,T)}+\frac{bA_T+aB_T}{\sqrt{-2a}(-a+2A_T^2)}\tanh \theta(0,T)+
  \left(A_T-\sqrt{\frac{-a}{2}}\right) \frac{{\lambda}{\mathcal M}}{ -a+2A_T^2}.\nonumber
  \end{eqnarray}

\bigskip

\section{3. APPLICATION TO THE BEHAVIORAL ECONOMICS}

The choice of the right-hand side of $ g $ seems artificial, but the example below shows that this form of control arises in real economic problems.

We assume that every investor solves the Merton problem of the portfolio selection \cite{Merton}.
Let us retell the statement of the problem, which each investor solves individually, following \cite{Oksendal}, Example 11.2.5.

The investor manages the portfolio $ V $, which consists of risky and risk-free assets, where $ h_1 $ and $ h_2 $ are  shares invested in risky and risk-free assets, respectively, $ h_1 + h_2 = 1 $. Let us set $ h_1 = h $, $ h_2 = 1-h $.
Prices for risky and risk-free assets satisfy
\begin{equation}
\label{(77)} dS_1 = \mu S_1dt + \sigma S_1dW_t,\quad dS_2 = r S_2dt,
\end{equation}
where $W_t$  is the standard Wiener process $\mu, r =\rm const$,
$\sigma={\rm const}
> 0$.
  Then
$\frac{dV}{V} = h_1\frac{dS_1}{S_1} +
h_2\frac{dS_2}{S_2}$, and
 the change in the value of the portfolio  is
\begin{equation}
\label{(V)} dV= (r + (\mu - r)h)V\,dt + \sigma h V\,  dW_t, \quad  V_t = v> 0.
\end{equation}
The investor wants to maximize the expected return on capital at some subsequent point of time $ T> t $ with respect to
 a utility function $ N (V) $, which is usually assumed to be increasing and convex upward.To get an explicit solution, the so-called HARA (hyperbolic absolute risk aversion) utility function is usually chosen\cite{Ingersoll}. Namely, $ N (v) = \frac{v^q}{q}$, $ q <1 $, $ q \ne 0 $, or $N(v)=\ln v$.
The latter function formally corresponds to the limit $ q \to 0 $.
In addition, the case $ q <0 $ corresponds to the strategy of the investor who prefers the least risky investments, $ q = 0 $ corresponds to risk-neutral strategies, $ q> 0 $ corresponds to the risk-prone investor \cite{BPS}, Sec.2.

The problem reduces to finding the function $ \Phi (t, v) $ and the control $ h_* = h _* (t, V) $, such  that
\begin{eqnarray}
\label{(11p)}\Phi(t,v)=\sup\limits_{h} \{J^h(t,v)\}=
J^{h_*}(t,v),\nonumber
\end{eqnarray}
where $h$ are all  admissible  Markov  control functions, $J^h(t,v)={\mathbb
E}^{t,v}[N(V_T^h)]$. In order to solve this problem we should define a differential operator
\begin{eqnarray}
\label{(13)} \mathcal{L}^h f = \frac{\partial f}{\partial t} + (\mu
h+r (1-h))v \,\frac{\partial f}{\partial v} + \frac{1}{2}\sigma^2
h^2 v^2 \,\frac{\partial^2 f}{\partial v^2}\nonumber
\end{eqnarray}
and solve the Hamilton-Jacobi-Bellman equation:
\begin{eqnarray}
\label{HJB0}\sup_h \{({\mathcal L}^h \Phi)(t,v)\}=0,\quad t\in(0,T),
\,v>0,\\
\Phi(T,v)=N(v),\quad \Phi(t,0)=N(0), \quad t<T.\nonumber
\end{eqnarray}
If $\partial_v\Phi>0$ and $\partial^2_{vv}\Phi<0$,  then the solution is
 $h(t,v)=\frac{(\mu-r)\partial_{v}\Phi}{v\sigma^2\partial^2
_{vv}\Phi}$. Substituting this expression in (\ref{HJB0}) gives
the following boundary value problem for  $\Phi$:
\begin{eqnarray}
\label{HJB1}
\partial_t\Phi+rv\partial_v\Phi-\frac{(\mu-r)\partial_v\Phi}{2\sigma^2\partial^2_{vv}\Phi}=0, \quad
t\in(0,T),
\,\,v>0,\\
\Phi(t,v)=N(v), \quad t=T \,\,\mbox{or}\,\, v=0.\nonumber
\end{eqnarray}
The solution (\ref{HJB1}) can be found in the form $\Phi(t,v) = \phi(t)N(v)$,
the corresponding optimal strategy at all $q<1$ is
\begin{eqnarray}\label{OS}
 h_* = \frac{(\mu - r)}{\sigma^2(1-q)}.
\end{eqnarray}
 From (\ref{(V)}) and $\rm It\hat{o}$'s formula we get
\begin{eqnarray}
\label{(22)} d \ln V = \Big[(\mu h + r(1-h)) - \frac{1}{2}\sigma^2
h^2 \Big]dt + \sigma h \,dW_t,\nonumber
\end{eqnarray}
therefore the capital  growth rate  $\displaystyle{\mathbb E}\frac{\ln
V}{t}$, that investor will receive guided by the optimal strategy (\ref{OS})
for all $q<1$ can be calculated as
\begin{eqnarray}
\label{(24)}{\mathbb E}\frac{\ln V}{t}= r +
\frac{(1-2q)(\mu-r)^2}{2\sigma^2(q-1)^2}.\nonumber
\end{eqnarray}

Now let us describe the collective behavior of investors.
We assume that they all manage the portfolio based on their own ideas about the parameter $\mu$. In other words, each fixed investor carries out
control based on the equation (\ref {(77)}) with its own choice of $ \mu $.
The "true" values of these parameters $\bar\mu$ is unknown to investors.
 These "true" values may differ from those accepted in the market, and they manifest themselves only in that the investor receives a penalty for their incorrect choice.

It is believed that the opinions of investors about the correct value of the drift parameter  $\mu$ are distributed along the entire axis, the maximum is initially at $\mu_0$.
Investors receive a penalty both for deviating from the "true" values of the drift and volatility parameters, and for deviating from the majority opinion.
A significant simplification that stems from the desire to obtain an analytical solution to the problem is the assumption that all investors treat risk the same way, guided by the same utility function $ N(v)$ and the same volatility $\sigma$.
During the control process, the initial distributions of $\mu$  change with the desire to maximize capital growth rate.

We obtain a typical  problem of the theory of mean-field games
(\ref{(1)}), when the random variable $\mu$
plays the role of $X$, subordinate to (\ref{X}), $K(X(T))= \frac{\ln
V(T)}{T}$, and
\begin{eqnarray}
g(t,X, m)=\beta\frac{\ln V(X)}{t} - \gamma (X - \bar{x})^2, 
\nonumber
\end{eqnarray}
$\beta\ge 0$, $\gamma\ge 0$,   $ \bar{x}=\rm const$,
The coefficients  $\beta$ and $\gamma$  can be considered equal to zero or not, depending on what problems we are studying.

We suppose that the volatility of a risky asset is known and an opinion is formed regarding its trend.

 The system of equations
(\ref{HJB})--(\ref{KFP}) has the following form:
\begin{eqnarray}
\label{(33)} &\partial_t\Phi + \frac{1}{2}(\partial_\mu\Phi)^2 +
\frac{\delta^2}{2}\partial^2_{\mu \mu}\Phi = -\beta (r +R(\mu -
r)^2) + \gamma(\mu - \bar{\mu})^2,
\\
&\partial_t m + \partial_{\mu}(m\partial_\mu\Phi) -
\frac{\delta^2}{2}\partial^2_{\mu \mu} m = 0,\nonumber
\\
&\Phi(\mu, T) = r + R(\mu - r)^2, \quad m(\mu, 0) =m_0(x),
 \label{(33.1)}
\end{eqnarray}
where $R= \frac{(1-2q)}{2\sigma^2(q-1)^2}$.

 In the notation of Sec.2
\begin{eqnarray}
&a=\beta R-\gamma,\quad b=2(\gamma\bar \mu-\beta Rr),\quad c=\beta r+\beta Rr^2-\gamma\bar \mu^2,\nonumber\\
&A_T=R,\quad B_T=-2Rr,\quad
C_T=r+Rr^2.\nonumber
\end{eqnarray}

According to the results of Sec.2, the solution of
(\ref{(33)})--(\ref{(33.1)}) exists for all $T>0$, if and only if $\beta R-\gamma<0$ and $R<\sqrt{\frac{\gamma-\beta R}{2}}$.
However, the expectation $\mathbb E (t)$ can always be determined. Namely, if $\beta R-\gamma<0$ ($a<0$), then for large $T$ the position of maximum  is close to  $$Q_*=\frac{r\beta
R-\gamma\bar \mu}{\beta R-\gamma}.$$ In other words, in this case, investors form an opinion about the asset. If $\beta
R-\gamma>0$ ($a>0$), then $\mathbb E (t)$ oscillates periodically, sometimes  deviating significantly from its average value. The frequency of these oscillations increases with  $a$. In this case, we say that investors cannot agree on the parameters of asset.

We refer to \cite{RN} for the analysis of this result from the point of view of the agent's behavior when investing for a long period of time.

\section{4. DISCUSSION}

 {\bf 1. Expectation versus maximum of the density}.
In \cite{RN} we consider a similar problem about tracing of the position of maximum on the Gaussian distribution density for the case of the pure diffusion ($\lambda=0$) and constant $a$ and $b$. It has been shown that the position of maximum is a solution to problem (\ref{EE}). It is quite natural result, since for the Gaussian distribution the position of maximum for the density coincides with the mathematical expectation.

Of course, one can set the problem about tracing of the position of maximum on the density in the present case of jump-diffusion. Nevertheless, here this problem is much more difficult, since it is impossible to follow the previous method and obtain an explicit solution: the probability density distribution, compatible with the jump-diffusion process, is expressed in special (Kummer) functions (see \cite{RSaakian}).

\bigskip

\noindent
 {\bf 2. Special form of $g(m,t,x)$}.
If the goal in the process of control is to minimize the distance from $\mathbb E(t)$, i.e. $(x-\mathbb E(t))^2$, such a result would be similar to the absence of any control. Indeed, if $g=-\gamma (x-\mathbb E(t))^2$, $\gamma$ is a positive constant, then we are in the situation of Corollary \ref{Cor}, and
$$g=-\gamma x^2 + 2\gamma \mathbb E(t) -  \gamma (\mathbb E(t))^2,\quad a=-\gamma,\, b_1= 2\gamma, \,b_2= b_0=0. $$
Thus, $\mathbb E''=0$ and the expectation changes linearly according to the final condition. The same result we would obtain if we set $\gamma=0$. This can be explained
by the fact that the presence of the term
$-\frac{\alpha^2}{2}$ already means that agents tend to resemble each other, this creates movement in the same direction.
The same effect has   $g(m)$ which
increases with $m$ (at least for the Gaussian distribution). This means that it is beneficial for agents to stay closer to the maximum of  $m$. However, as has been shown numerically, the concentration near maximum of distribution becomes faster if the terms of such kind presents.

\bigskip
\noindent
{\bf 3. Related works}.
The possibility of reducing the MFG system to the matrix Riccati equations has been known for other models (e.g.  \cite{Bensoussan}, \cite{Fatone}, \cite{Trusov}, \cite{Yong}). In \cite{TradeCrowding} the authors obtain the equation for the mathematical expectation and use it for the analysis of a model of trading. This equation also is a second order ODE with constant coefficients. In \cite{Graber}, \cite{Lucas}, \cite{Poretta} models with underlying jump-diffusion were considered.

\section{ACKNOWLEDGMENTS}
O.R. was partially supported by  Ultra Quantum Inc. and the Moscow Center for Fundamental and Applied Mathematics. Special thanks to Prof. Alex Kurganov for preliminary numerical experiments.

\end{document}